# An attention-based deep learning network for predicting Platinum resistance in ovarian cancer

Haoming Zhuang[1], Beibei Li[2], Jingtong Ma[1], Patrice Monkam[1], Shouliang Qi[1], Wei Qian[1], Dianning He[1*]

[1]College of Medicine and Biological Information Engineering, Northeastern University, Shenyang, China

[2] Department of Radiology, Shengjing Hospital of China Medical University, Shenyang, China

*Correspondence to: Dianning He (email: hedn@bmie.neu.edu.cn)

# Abstract

**Background:** Ovarian cancer is among the three most frequent gynecologic cancers globally. High-grade serous ovarian cancer (HGSOC) is the most common and aggressive histological type. Guided treatment for HGSOC typically involves platinum-based combination chemotherapy, necessitating an assessment of whether the patient is platinum-resistant. The purpose of this study is to propose a deep learning-based method to determine whether a patient is platinum-resistant using multimodal positron emission tomography/computed tomography (PET/CT) images.

**Methods:** 289 patients with HGSOC were included in this study. An end-to-end SE-SPP-DenseNet model was built by adding Squeeze-Excitation Block (SE Block) and Spatial Pyramid Pooling Layer (SPPLayer) to Dense Convolutional Network (DenseNet). Multimodal data from PET/CT images of the regions of interest (ROI) were used to predict platinum resistance in patients.

**Results:** Through five-fold cross-validation, SE-SPP-DenseNet achieved a high accuracy rate and an area under the curve (AUC) in predicting platinum resistance in patients, which were 92.6% and 0.93, respectively. The importance of incorporating SE Block and SPPLayer into the deep learning model, and considering multimodal data was substantiated by carrying out ablation studies and experiments with single modality data.

**Conclusions:** The obtained classification results indicate that our proposed deep learning framework performs better in predicting platinum resistance in patients, which can help gynecologists make better treatment decisions.

# Introduction

Ovarian cancer is the third most common gynecologic cancer in the world, with approximately 313,959 new cases and 207,252 deaths, ranking eighth in incidence and mortality among female cancers worldwide [1]. There are five main histologic types of ovarian cancer, of which high-grade serous ovarian cancer (HGSOC) is the most common and aggressive histologic type of ovarian cancer [2]. The guideline treatment for HGSOC, as suggested by the National Comprehensive Cancer Network. Consists of the removal of the tumor followed by platinum-based combination chemotherapy [3]. However, because some patients are resistant to chemotherapy, they are at high risk of recurrence and require further treatment. In general, the indicator of whether a patient is resistant or sensitive to subsequent platinum-based chemotherapy is based on the length of the platinum-free interval (PFI), which is defined as the time interval between completion of platinum-based combination chemotherapy and disease progression [4]. After initial treatment, patients are considered "platinum-sensitive" if they relapse after 6 months or more, and "platinum-resistant" if they relapse within 6 months. Platinum-resistant patients typically have a low response rate to subsequent chemotherapy (<15%), with a progression-free survival of 3-4 months and a median survival of <12 months [5].

Regardless of the great advancements in precision medicine, proactively and accurately predicting whether a patient is platinum-resistant remains a challenge. If a patient is likely to be platinum-resistant, it is possible to treat the patient more effectively than the standard of care of platinum-based combination chemotherapy. For example, the approach and timing of surgery can be optimized for secondary cytoreductive surgery, thereby limiting the development potential of the drug-resistant subclonal tumor population [6]. At the same time, drug-resistant patients can be tested more frequently to detect tumor recurrence without delay. In addition, platinum resistance is a simple indicator of sensitivity to poly (ADP-ribose) polymerase inhibitors (PARPi) [7]. Hence, accurately predicting platinum resistance in patients will reduce the need for unnecessary and cumbersome clinical testing. Therefore, if platinum-resistant patients can be accurately predicted, they will be able to take full advantage of the benefits of precision medicine.

Currently, most methods for predicting platinum-resistant patients use biomarkers, expression of specific genes or proteins, tumor immunohistochemistry, etc. Kuhlmann et al predicted platinum resistance in ovarian cancer patients by using excision repair cross-complementation group 1 (ERCC1) positive circulating tumor cells as a predictive biomarker for platinum resistance [8]. Wu et al found that the risk of platinum resistance was 60-fold higher in patients with high co-expression levels of glutathione peroxidase (GPX4) and cystine/glutamate antiporter SLC7A11 than in those with low co-expression levels [9]. They could determine the platinum resistance by assessing whether patients had high co-expression of GPX4 and SLC7A11. However, these methods are associated with high assay costs, invasiveness, and additional time delays. In contrast, in this study, we adopt 18F-2-fluoro-2-deoxy-D-glucose positron emission tomography/computed tomography (18F-FDG PET/CT), which can non-invasively and more effectively obtain basic information about the tumor, such as the size and location of the primary lesion and metastatic lesions, and use the images to determine whether the patient has platinum resistance. This eliminates the labor-intensive process of manual extraction and detection required by traditional methods, as well as the impact of operator error.

Herein, we propose a deep learning-based approach to predict the presence of platinum resistance in patients using their PET/CT medical images. Our method offers improved detection efficiency compared to conventional approaches as it follows an end-to-end workflow. The proposed deep model prioritizes vital information for classification and multilevel pooling by integrating Squeeze-Excite Block (SE Block) and Spatial Pyramid Pooling Layer (SPPLayer) to improve the classification accuracy. Ultimately, the Dense Convolutional Network (DenseNet) model with SE Block and SPPLayer (SE-SPP-DenseNet) demonstrated the highest precision in predicting platinum resistance in patients.

## Method

### Patients

This retrospective study was approved by the Radiology Review Committee of Sheng Jing Hospital of China Medical University and adhered to the principles and requirements of the Declaration of Helsinki. The study involved retrospective analysis of prospectively collected data on 289 patients with high-grade plasma ovarian cancer between January 2013 and December 2017 at the hospital.

**PET/CT data**

Patients who had fasted and refrained from food and drink for more than 6 hours were administered a one-hour intravenous infusion of 161-361 MBq of $^{18}$F-FDG before undergoing PET/CT scanning (GE Discovery; GE Healthcare, Inc. Milwaukee, WI). A 3D PET model with a matrix size of 512 × 512 and an exposure time of 2 minutes per bed was utilized. Following attenuation correction with CT (120 kV, 80 mA), PET images were reconstructed through a time-of-flight and point-spread function algorithm that incorporated 2 iterations and 20 subsets.

**Data preparation**

In our study, the physicians delineated the regions of interest (ROI) in the PET and CT images of the patients. These image data were imported into MATLAB 2021b (MathWorks, Natick, MA, USA) and the ROIs were extracted, and then stacked into two-dimensional data for each channel. The resulting data was used as inputs for the deep learning network. The data preparation process is shown in Figure 1.

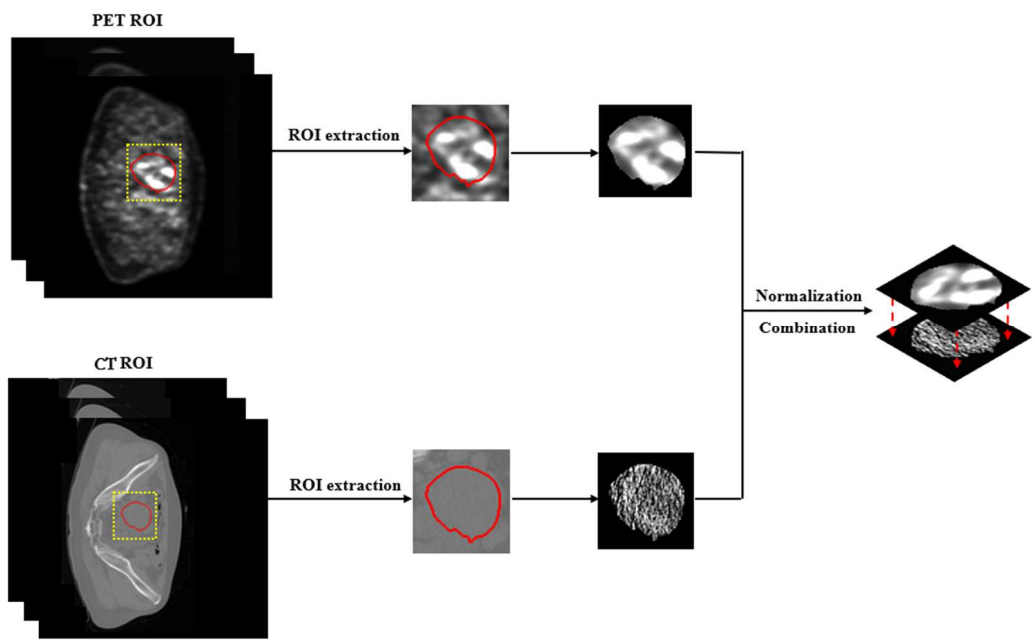

Fig. 1 Data preparation process

**Deep Learning Models for Automatic Platinum Resistance Detection**

The deep learning models used in this study were implemented using the Pytorch framework and Python 3.10, and were trained on a single workstation with an Intel Core i9-13900H processor, NVIDIA 4060 graphics card, and 8 GB of RAM. In our study, we constructed a classification model utilizing SE-SPP-DenseNet models to ascertain the presence of platinum resistance through analysis of patients imaging data. A significant difference in the number of platinum-resistant patients (97) and platinum-sensitive patients (192) would have resulted in overfitting of the model. Thus, data augmentation was executed on the platinum-resistant patient data. Then, the data was rotated by 90 degrees to double its volume, ensuring relative balance with the data of platinum-sensitive patients. The dataset was allocated on a per-patient basis in the ratio of training set: validation set: test set = 0.8:0.1:0.1.

We utilized Dense Convolutional Network (DenseNet), whose structure is shown in Figure 2 [11].

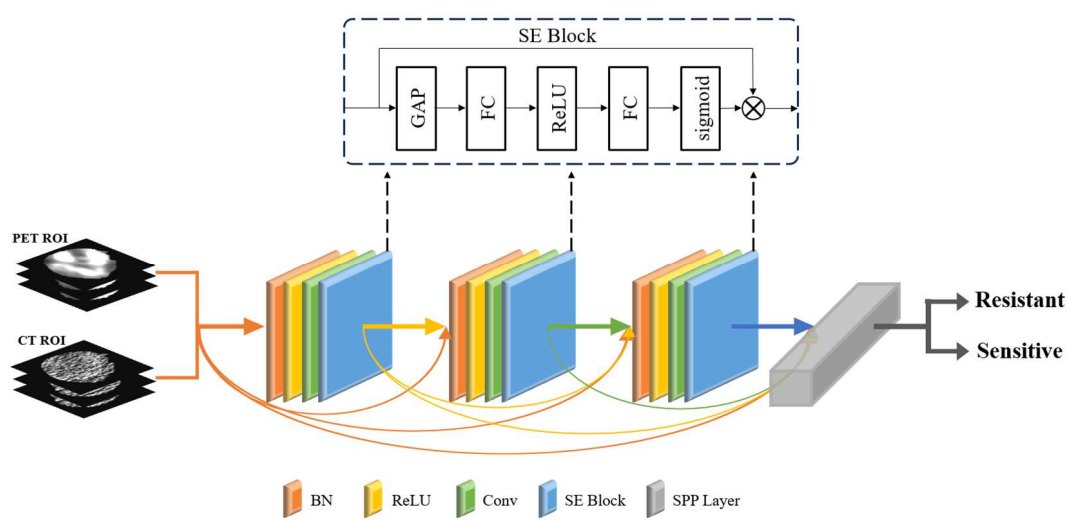

**Fig. 2** Structure of DenseNet

There are four variants of the DenseNet models: DenseNet-121, DenseNet-169, DenseNet-201, and DenseNet-264. For this study, we use DenseNet-121 as the backbone of the convolutional neural network (CNN) architecture, and we add squeeze-excitation block (SE Block) [12] and Spatial Pyramid Pooling Layer (SPPLayer) for prediction. The SE Block's structure is shown in Figure 2, where a four-step operation is performed between the inputs and outputs of the convolutional layers in the network.

We incorporated the SPPLayer into the network which replaced the subsequent pooling layer in the DenseNet output. This enables the extraction of multi-scale features and inputting them into the fully connected layer with a fixed size, regardless of the image's input size, after selecting the suitable spatial bins. It is worth noting that generated experimental images were resized to 224×224.

We also utilized the more common deep residual network18 (ResNet18) for medical image classification [13]. ResNet18 model increases the number of layers of the network to extract richer features at different depths. At the same time, ResNet18 prevents gradient dispersion or gradient explosion by regularizing the initial and intermediate regularization layers. ResNet18's network structure is shown in Figure 3. In this study, we also added SE Block and SPPLayer to ResNet18.

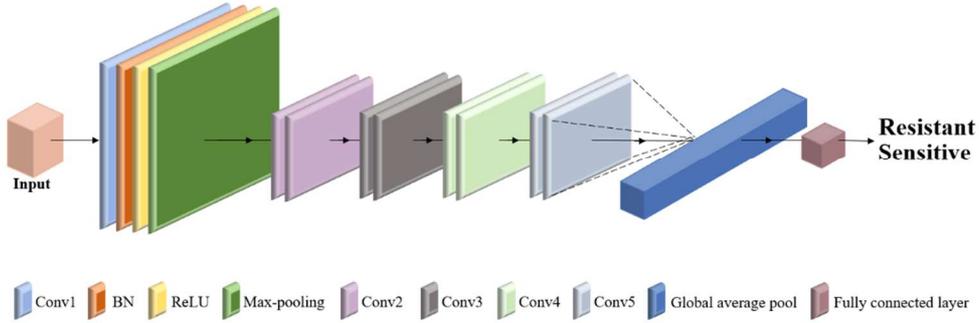

**Fig. 3** Resnet18 network structure

Furthermore, we used Swin Transformer model for classification, which employs a hierarchical construction approach and its structure is shown in Figure 4 [14].

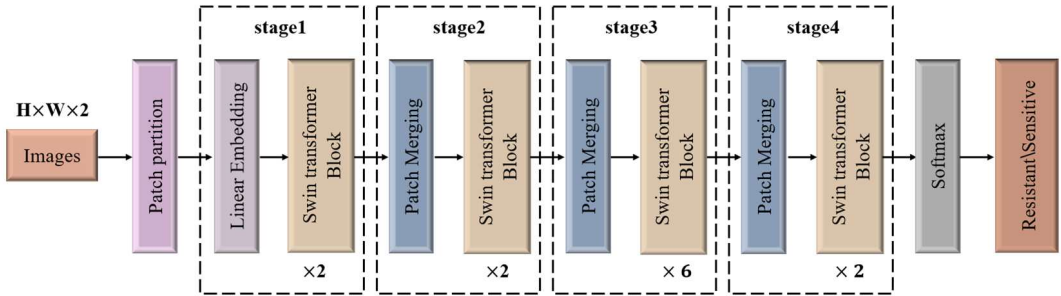

**Fig. 4** Structure of Swin Transformer

In our study the Patch Partition Block down samples the image 4 times, 8 times and 16 times respectively, which helps to construct the task of target classification based on this. The Linear Embedding Layer performs a linear variation of our input two-channel image, and the Patch Merging Layer performs a depth doubling and halving of the height and width of the feature maps after the chunking of the image. The Swin Transformer Block is the key part of the Swin Transformer model, which consists of two structures. The only difference between them is that one uses Windowed Multi-Head Self-Attention (W-MSA) and the other uses Shifted Windowed Multi-Head Self-Attention (SW-MSA).

**Quantitative assessment metrics**

In this study, the accuracy, sensitivity, specificity, positive predictive value (PPV), and negative predictive value (NPV) of the above three deep learning models for predicting platinum resistance of patients in the test set were calculated, the receiver operating characteristic (ROC) curves and the confusion matrixes were constructed, and the area under the curve (AUC) was calculated.

# Results

### Patient Characteristics

289 patients were eligible and included in the study, of which the number of platinum-resistant patients was 97 and the number of platinum non-resistant patients was 192. And after data preprocessing and augmentation, the image data of platinum-resistant patients were about 4158 images, and the image data of platinum non-resistant patients were about 4534 images.

### Parameter settings for each model

After comparing the results of several preliminary experiments, the specific hyperparameters of DenseNet, ResNet18 and Swin Transformer are set to the following parameters as shown in Table 1. The learning rate is set to 0.1 times of the original if the loss of the validation set does not decrease after 10 epochs during the training process. And in the process of pre-experimentation, we found that all three models can converge before the Epoch number is equal to 50, so we used a fixed Epoch number equal to 50, and set the Early Stopping strategy to stop training if the loss of the validation set does not decrease after 20 epochs, so as to prevent the overfitting phenomenon from occurring. Meanwhile, SGD and Binary Cross Entropy Loss (BCE Loss) are selected as the network optimizer and loss function, respectively. During the training period, we also perform image augmentation, including inversion of the vertical direction of the image and normalization of the image by calculating the mean and variance of the image. However, it is important mentioning that the image augmentation is only performed for the images during the training period, and image augmentation is not performed for the images used for validation and testing.

**Table. 1** Parameters for model training

| Parameters | Epoch | Learning Rate | Batch Size | Momentum | Weight Decay | Optimizer |
|---|---|---|---|---|---|---|
| ResNet18 | 50 | 0.01 | 64 | 0.9 | 0.0001 | SGD |
| DenseNet | 50 | 0.01 | 48 | 0.9 | 0.0001 | SGD |
| Swin Transformer | 50 | 0.001 | 64 | 0.9 | 0.0001 | SGD |

**Platinum resistance classification results**

In this study, a detection framework based on deep learning techniques was proposed to predict whether a HGSOC patient is platinum resistant or not using multimodal images from PET/CT. We investigated the performance of three different deep learning models including DenseNet, ResNet18 and Swin Transformer.

Three models were trained and tested using five-fold cross-validation. The SE-SPP-DenseNet model achieved an accuracy of 92.6%, sensitivity of 86.3%, specificity of 96.1%, PPV of 95.7%, and NPV of 85.7%. The quantitative evaluation of the ResNet18 model, adding SE Block and SPPLayer (SE-SPP-ResNet18), showed an accuracy of 88.2%, sensitivity of 78.2%, specificity of 95.2%, PPV of 95.8% and NPV of 75.1%. The Swin Transformer model obtained an accuracy of 83.1%, sensitivity of 85.7%, specificity of 77.5%, PPV of 84.2%, and NPV of 79.5%.

The ROC curves for the three models mentioned above are shown in Figure 5, where the AUC of the SE-SPP-DenseNet, SE-SPP-ResNet18 and Swin Transformer models are 0.93, 0.92, and 0.87, respectively.

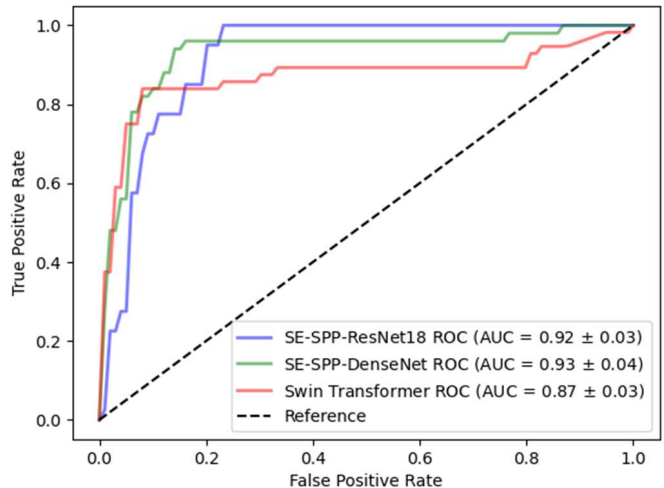

**Fig. 5** ROC curves of SE-SPP-DenseNet, SE-SPP-ResNet18 and Swin Transformer

**Effectiveness of SE block and SPPLayer module for Platinum resistance assessment**

We also performed some ablation studies to further validate the effectiveness of SE Block and SPPLayer in improving the prediction performance of the DenseNet and ResNet18 models. The results of the ablation studies are shown in Table 2.

Table. 2 Classification results of different models

|  | Accuracy | Sensitivity | Specificity | PPV | NPV |
|---|---|---|---|---|---|
| Resnet18 | 77.8% | 68.2% | 78.3% | 76.2% | 66.7% |
| Resnet18 + SE Block | 82.2% | 68.6% | 87.5% | 89.8% | 64.5% |
| Resnet18 + SPPLayer | 84.2% | 80.4% | 85.1% | 85.7% | 77.6% |
| **Resnet18 + SE Block + SPPLayer** | **88.2%** | **78.2%** | **95.2%** | **95.8%** | **75.1%** |
| DenseNet | 74.3% | 73.2% | 75.4% | 74.5% | 73.4% |
| DenseNet + SE Block | 77.2% | 76.2% | 78.5% | 77.4% | 75.1% |
| DenseNet + SPPLayer | 83.3% | 80.4% | 87.5% | 88.7% | 76.1% |
| **DenseNet + SE Block + SPPLayer** | **92.6%** | **86.3%** | **96.1%** | **95.7%** | **85.7%** |
| Swin Transformer | 83.1% | 85.7% | 77.5% | 84.2% | 79.5% |

**The Importance of PET Images**

Furthermore, this study uses PET and CT images as multimodal data for training and testing, but many patients are currently diagnosed with CT only, not PET/CT. To illustrate the importance of PET images in predicting platinum resistance, we performed experiments on the three models above using only CT images to train and test. The AUC of the SE-SPP-DenseNet, SE-SPP-ResNet18 and Swin Transformer models are 0.87, 0.84, and 0.64, respectively. The ROC curves and quantitative results for the three models above are shown in Figure 6 and Table 3.

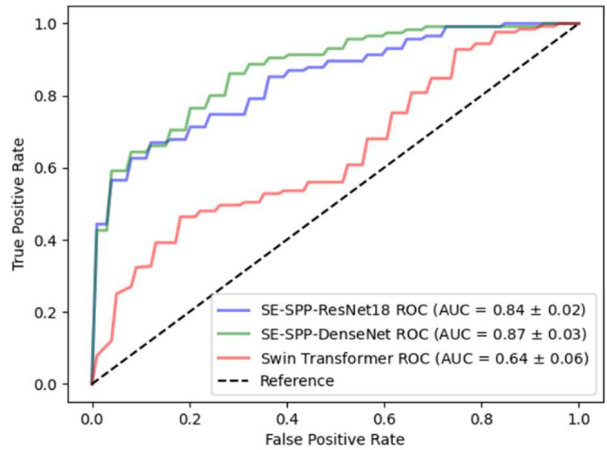

Fig. 6 ROC curves of ResNet18, DenseNet and Swin Transformer

**Table. 3** Classification results of different models

|  |  | Accuracy | Sensitivity | Specificity | PPV | NPV |
|---|---|---|---|---|---|---|
| Resnet18 + SE Block + SPPLayer | SM | 69.6% | 97.6% | 39.1% | 63.7% | 95.0% |
|  | **MM** | **88.2%** | **78.2%** | **95.2%** | **95.8%** | **75.1%** |
| DenseNet + SE Block + SPPLayer | SM | 76.7% | 87.2% | 65.2% | 73.4% | 82.0% |
|  | **MM** | **92.6%** | **86.3%** | **96.1%** | **95.7%** | **85.7%** |
| Swin Transformer | SM | 61.7% | 77.6% | 33.9% | 56.3% | 56.8% |
|  | **MM** | **83.1%** | **85.7%** | **77.5%** | **84.2%** | **79.5%** |

*SM* single modality, *MM* multimodality

# Discussions

In this study, we developed deep learning models to predict platinum resistance in HGSOC patients. PET/CT images of patients' ovarian cancer lesions were used as two-channel multimodal inputs for prediction using an improved DenseNet model. We improved the accuracy of predicting platinum resistance by adding SE Block and SPPLayer to DenseNet model. The SE-SPP-DenseNet model achieved the highest accuracy of 92.6% in predicting platinum resistance.

Based on Figure 5 and Table 2, it is evident that the predictive accuracy of DenseNet surpasses that of ResNet. This can be attributed to each DenseNet layer acquiring supplementary inputs from all previous layer connections, in addition to integrating its own outputs with all succeeding layers. DenseNet is different from ResNet in that it does not simply add the feature outputs of earlier layers to the inputs of later layers, but connects these inputs and outputs using fewer parameters. Through the dense connectivity between the layers, better classification results can be achieved. Additionally, the selection of the network should be determined by the data complexity. This study utilized ResNet18 as a model, while ResNet152, although speculated to yield superior results, exhibited unsatisfactory performance during training iterations. Increasing model complexity may improve fitting but reduce generalization ability, leading to overfitting especially in case of limited training samples. The optimal solution used by the Stochastic Gradient Descent (SGD) algorithm during training is not a global optimal solution, but a local optimal solution, so the more complex the network, the more complex the solution space, resulting in the inability to obtain an optimal solution using SGD.

With the increasing complexity of neural network learning, the amount of stored information is increasing, so it is more advantageous to focus resources on the critical inputs for the current classification task if computing power is constrained. Therefore, we added the SE Block, which reduces the focus on other information, and as can also be seen from Table 2, the addition improves the accuracy of the model in terms of predicting the patient's platinum resistance. Furthermore, due to the non-uniformity of ROI sizes, it is necessary to adjust the various ROI input sizes for training and testing according to the distribution of ROI sizes. The test results can then be used to determine the appropriate ROI size for scaling up or down. SPPLayer not only enables the network to adapt to different sizes of input data, but also handles object distortion and spatial layout variations through multi-level pooling, thus improving the classification accuracy [16]. Therefore, SPPLayer was added to the model. As can be seen in Table 2, the model exhibits an improvement in the quantitative assessment index of platinum resistance in patients after the addition of SPPLayer.

Compared to currently available learning models for predicting platinum resistance in patients, our current study has a different predictive approach and further improvements. Hwangbo et al developed a machine learning model to predict platinum resistance in patients by using four machine learning algorithms: logistic regression (LR), random forest (RF), support vector machine (SVM), developed a machine learning model to predict platinum resistance in patients [15]. In this study, SE-SPP-DenseNet model was established to predict platinum resistance in patients using deep learning, which has better model generalization ability compared to machine learning, and can extract and learn features directly from the raw data, which simplifies the need to collect relevant features related to platinum resistance in the process of machine learning. It greatly reduces the unnecessary work in the process of medical imaging applications. Wu et al. were able to predict the risk of platinum resistance in patients based on the level of protein expression, but they were still unable to accurately identify the patient [9]. Therefore, the development of a deep learning model based on PET/CT images to predict platinum resistance in patients with HGSOC can help physicians make the best treatment decisions.

Our study has several limitations. Firstly, as this is a retrospective study, only patient cases within the four-year timeframe were selected as experimental data, potentially introducing bias in the results. Secondly, the study's applicability is limited due to a small sample size, thus requiring an increase in data collection for future research.

## Conclusion

In this study, we proposed a method using the deep learning model SE-SPP-DenseNet to predict platinum resistance in patients based on PET/CT multimodal images of ROI. Ablation studies demonstrate that adding SE Block and SPPLayer to DenseNet can enhance the classification effect of the original model. Experiments utilizing single modality CT images demonstrated the importance of multimodal imaging for platinum resistance prediction. The SE-SPP-DenseNet model accurately predicts platinum resistance in patients, with a final accuracy of 92.6% and an AUC of 0.93. We believe that gynecologists can use this model for assisted diagnosis to determine the best treatment plan and prognostic measures based on whether a patient has platinum resistance.